\documentclass[12pt]{mcom-l}
\usepackage{amsmath,graphicx,psfrag}

\newcommand{\ii}{\mathrm{i}}
\DeclareMathOperator{\cs}{cs}
\DeclareMathOperator{\e}{e}
\DeclareMathOperator{\T}{T}
\DeclareMathOperator{\sech}{sech}
\numberwithin{equation}{section}

\begin{document}

\title{Maa{\ss} cusp forms for large eigenvalues}
\author{Holger Then}
\address{Abteilung Theoretische Physik, Universit\"{a}t Ulm, 89069 Ulm, Germany}
\email{holger.then@physik.uni-ulm.de}
\urladdr{http://www.physik.uni-ulm.de/theo/qc/group.html}
\thanks{The encouraging advice of Prof. Dennis A. Hejhal and Prof. Frank Steiner is gratefully acknowledged. Best thanks are also due to Ralf Aurich, Jens Bolte, David Farmer and Andreas Str\"{o}mbergsson. For the stay in Prof. Hejhal's group, where part of the work was done, the author was supported by the European Commission Research Training Network HPRN-CT-2000-00103. Currently, the author is supported by the Deutsche Forschungsgemeinschaft under the contract no. DFG Ste 241/16-1. The computations were run on the computers of the Universit\"{a}ts-Rechenzentrum Ulm.}


\date{November 26, 2002}


\begin{abstract} We investigate the numerical computation of Maa{\ss} cusp forms for the modular group corresponding to large eigenvalues. We present Fourier coefficients of two cusp forms whose eigenvalues exceed $r=40000$. These eigenvalues are the largest that have so far been found in the case of the modular group. They are larger than the $130$millionth eigenvalue. \end{abstract}

\maketitle

\section{Introduction} To extend the classical theory of Dirichlet series with Euler products, Maa{\ss} \cite{Maass1949} studied non-analytic automorphic functions, nowadays called Maa{\ss} waveforms. They are defined in the upper half-plane,
\begin{align*} {\mathcal H}=\{z=x+\ii y; \quad x,y \text{ real}, \quad y>0\}, \end{align*}
equipped with the hyperbolic metric
\begin{align*} ds^2=\frac{dx^2+dy^2}{y^2}. \end{align*}
Maa{\ss} waveforms $f$ are real analytic eigenfunctions of the hyperbolic Laplacian,
\begin{align} \label{EigenvalueEquation} (\Delta+\lambda)f(z)=0. \end{align}
The Laplacian in the hyperbolic metric reads
\begin{align*} \Delta=y^2(\frac{\partial^2}{\partial x^2}+\frac{\partial^2}{\partial y^2}), \end{align*}
and is invariant under the group of linear fractional transformations
\begin{align*} z\mapsto\gamma z=\frac{az+b}{cz+d};\quad a,b,c,d \text{ real},\quad ad-bc=1.\end{align*}
This group is isomorphic to the group of matrices
\begin{align*} \gamma=\begin{pmatrix} a&b\\c&d \end{pmatrix} \in\operatorname{SL}(2,\mathbb{R})/\{\pm1\}. \end{align*}
In addition, Maa{\ss} waveforms are required to satisfy the automorphy condition
\begin{align} \label{AutomorphyCondition} f(\gamma z)=f(z) \quad \forall\gamma\in\Gamma \end{align}
relative to a cofinite discrete subgroup
\begin{align*} \Gamma\subset\operatorname{SL}(2,\mathbb{R})/\{\pm1\}, \end{align*}
and to satisfy the bound
\begin{align} \label{Bounds} f(z)=O(y^{\kappa}) \quad \text{for } y\to\infty, \end{align}
uniformly in $x$ for some positive constant $\kappa$, and similarly in the other cusps. Maa{\ss} waveforms which vanish in all the cusps, i.e., for which
\begin{align*} f(z)\to0 \quad \text{as} \quad \Im z\to+\infty \end{align*}
and analogously at the other cusps are called Maa{\ss} cusp forms. For references, cf. e.g. \cite{Selberg1956,Roelcke1966,Hejhal1983,Terras1985,Miyake1989,Venkov1990,Iwaniec1995}. \par
We choose the discrete group $\Gamma$ to be the modular group,
\begin{align*} \Gamma=\operatorname{PSL}(2,\mathbb{Z})=\operatorname{SL}(2,\mathbb{Z})/\{\pm1\}. \end{align*}
It is generated by two elements
\begin{align*} \begin{pmatrix} 1&1\\0&1 \end{pmatrix},\quad \begin{pmatrix} 0&-1\\1&0 \end{pmatrix}, \end{align*}
which are isomorphic to the translation and the inversion
\begin{align*} z\mapsto z+1, \quad z\mapsto -\frac{1}{z}, \end{align*}
and has a fundamental domain which can be chosen to be
\begin{align*} {\mathcal F}=\Gamma\backslash{\mathcal H}=\{z=x+\ii y\in{\mathcal H}; \quad |x|<\frac{1}{2}, \quad |z|>1\}. \end{align*}
Maa{\ss} cusp forms are square integrable over the fundamental domain
\begin{align*} \int_{\mathcal F} |f(z)|^2 \, d\mu<\infty, \end{align*}
where the volume element is
\begin{align*} d\mu=\frac{dx \, dy}{y^2}. \end{align*}
The reflection symmetry of the fundamental domain $\mathcal F$ in the line $x=0$ implies that the Maa{\ss} waveforms can be chosen such that they fall into two symmetry classes, the even functions $f(x+\ii y)=f(-x+\ii y)$, and the odd ones $f(x+\ii y)=-f(-x+\ii y)$, respectively. From the definition of Maa{\ss} waveforms (\ref{EigenvalueEquation}), (\ref{AutomorphyCondition}), and (\ref{Bounds}), it follows that they can be expanded into Fourier series,
\begin{align} \label{FourierExpansion} f(z)= u_0(y) + \sum_{n\in\mathbb{N}} a_n y^{\frac{1}{2}} K_{\ii r}(2\pi n y) \cs(2\pi n x), \end{align}
where
\begin{align*} u_0(y)=\begin{cases} b_0 y^{\frac{1}{2}+\ii r} + b_1 y^{\frac{1}{2}-\ii r} & \text{if } r\not=0, \\ b_2 y^{\frac{1}{2}} + b_3 y^{\frac{1}{2}} \ln y & \text{if } r=0, \end{cases} \end{align*}
and
\begin{align*} \cs(x)=\begin{cases} 2\cos(x) & \text{for the even Maa{\ss} waveforms}, \\ 2\sin(x) & \text{for the odd ones}. \end{cases} \end{align*}
$K_{\ii r}(x)$ is the K-Bessel function (see appendix \ref{KBessel}) whose order is connected with the eigenvalue $\lambda$ by
\begin{align*} \lambda=r^2+\frac{1}{4}. \end{align*}
While keeping in mind that $\lambda$ is the true eigenvalue, we will often call $r$ to be the eigenvalue instead. \par
According to the Roelcke-Selberg spectral resolution of the Laplacian \cite{Selberg1956,Roelcke1966}, its spectrum contains both a discrete and a continuous part. The discrete part of the spectrum is spanned by the constant eigenfunction $f_0$ and a countable number of Maa{\ss} cusp forms $f_1,f_2,f_3,\ldots$ which we take to be ordered with increasing eigenvalues $0=\lambda_0<\lambda_1\le \lambda_2\le \lambda_3\le \ldots$. The continuous part of the spectrum $\lambda=r^2+\frac{1}{4}\ge\frac{1}{4}$ is spanned by the Eisenstein series $E(z,\frac{1}{2}+\ii r)$ which are known analytically \cite{Maass1949,Kubota1973}. The functions $\Lambda(\frac{1}{2}+\ii r)E(z,\frac{1}{2}+\ii r)$ are even and their Fourier coefficients are given by
\begin{align*} b_0=\Lambda(\frac{1}{2}+\ii r), \quad b_1=\Lambda(\frac{1}{2}-\ii r), \quad a_n=\sum_{\substack{c,d\in\mathbb{Z}\\cd=n}} \left| \frac{c}{d} \right|^{\ii r}, \end{align*}
with
\begin{align*} \Lambda(s)=\pi^{-s}\Gamma(s)\zeta(2s). \end{align*}
The positive eigenvalues and their associated Maa{\ss} cusp forms are not known analytically. Hence one has to calculate them numerically. \par
References concerning this computational work in the case of the modular group are: \cite{Cartier1971,AtkinSwinnerton1971,Haas1977,Cartier1978,Hejhal1981,GolovcanskiiSmotrov1982,HejhalBerg1982,GolovcanskiiSmotrov1984,Stark1984,Winkler1988,Hejhal1991,CsordasGrahamSzepfalusy1991,Huntebrinker1991,Schmit1991,HejhalRackner1992,Hejhal1992a,Hejhal1992b,Steil1992,HejhalArno1993,Steil1994,Hejhal1999,Avelin2003}. The first breakthrough to go beyond $r=27.284$ was made by Hejhal \cite{Hejhal1991} who computed the first $123$ eigenvalues and $36$ more in three intervals around $r\approx125$, $r\approx250$, and $r\approx500$, respectively. He used the truncated Fourier expansion
\begin{align} \label{TruncatedFourierExpansion} f(z)=\sum_{n=1}^{M} a_n y^{\frac{1}{2}} K_{\ii r}(2\pi n y) \cs(2\pi n x) + [[\varepsilon]] \end{align}
in the automorphy condition
\begin{align*} f(z)=f(-\frac{1}{z}) \end{align*}
and obtained a linear system of equations,
\begin{align*} \sum_{n=2}^{M} a_n I_n(z_m)=-I_1(z_m), \quad z_m\in{\mathcal F}, \quad 1\le m\le M-1, \end{align*}
with
\begin{multline*} I_n(z)=y^{\frac{1}{2}} K_{\ii r}(2\pi n y) \cs(2\pi n x) \\ - (\Im(-\tfrac{1}{z}))^{\frac{1}{2}} K_{\ii r}(2\pi n\Im(-\tfrac{1}{z})) \cs(2\pi n\Re(-\tfrac{1}{z})), \end{multline*}
where, for suitable $M$, the error term $[[\varepsilon]]$ is of negligible size and can be omitted. After multiplication of this linear system of equations with $\e^{\frac{\pi r}{2}}$, it was solved for successive $r$ values on a grid. Eigenvalues were found by checking whether the coefficients are multiplicative,
\begin{align} \label{MultiplicativeRelations} a_1=1, \quad a_{mp}=a_m a_p - a_{\frac{m}{p}}, \quad p \text{ prime}, \end{align}
with the convention $a_{\frac{m}{p}}=0$ if $p$ does not divide $m$. Because $I_n(z_m)$ gets small for large $n$, the linear system of equations is unstable. \par
An attempt to get around these instabilities was carried out by Stark \cite{Stark1984}, Hejhal and Arno \cite{HejhalArno1993}, and Steil \cite{Steil1992,Steil1994}. They used the eigenvalue equations
\begin{align*} \T_m f(z)=t_m f(z) \end{align*}
of the Hecke operators
\begin{align*} \T_m f(z)=\frac{1}{\sqrt{m}} \sum_{\substack{ad=m \\ b \text{ mod } d,\ d>0}} f(\frac{az+b}{d}); \end{align*}
see Maa{\ss} \cite{Maass1949,Maass1964}. The Hecke operators $\T_m$ are self-adjoint, commute with the Laplacian, with the symmetry of the fundamental domain, and amongst each other. They are multiplicative
\begin{align*} \T_m\T_n f(z)=\sum_{\substack{d|(m,n) \\ d>0}} \T_{\frac{mn}{d^2}} f(z), \end{align*}
and their eigenvalues are connected with the Fourier coefficients of the Maa{\ss} cusp forms by
\begin{align*} a_m=a_1 t_m. \end{align*}
Normalizing Maa{\ss} cusp forms according to $a_1=1$, the non-linear system of equations
\begin{gather*} \T_p f(z)=a_p f(z) \\ a_{mp}=a_m a_p - a_{\frac{m}{p}}, \quad p \text{ prime}, \end{gather*}
allowed Steil to compute all eigenvalues up to $r=350$ ($4401$ even and $4776$ odd eigenfunctions) and between $r=500$ and $510$ ($395$ even and $410$ odd). He was also able to compute eigenvalues around $r\approx4000$. \par
Finally, Hejhal \cite{Hejhal1999} found a linear stable algorithm for computing Maa{\ss} cusp forms together with their eigenvalues. His algorithm is based on finite Fourier transforms and implicit automorphy, and can be applied to holomorphic cusp forms as well as to Maa{\ss} cusp forms. Furthermore, his algorithm can also be applied to non-arithmetic groups and can be extended to groups whose fundamental domain has several cusps \cite{SelanderStrombergsson2002}. With this algorithm, Hejhal found eigenvalues around $r\approx11000$. The main obstacle to go beyond was a lack of further memory. Our goal in the present paper will be to obtain larger eigenvalues. We keep the main ideas, but optimize the algorithmic procedure used in finding the eigenvalues. Furthermore, we make careful use of the memory. This enables us to compute eigenvalues up to $r\approx40000$. Limitations to go beyond this are due not to lack of memory, but, rather to CPU time. The latter (which scales with the third power of $r$) exceeds four weeks on a $750$ MHz SUN UltraSPARC-III processor. Currently, the only ``larger'' Maa{\ss} cusp forms available on the numerical front are those explored by Hejhal and Str\"{o}mbergsson \cite{HejhalStrombergsson2001} in their recent work with waveforms of CM-type, i.e., waveforms on congruence subgroups which arise as lifts of automorphic forms on $\operatorname{GL}(1)$.

\section{Hejhal's algorithm} We make use of Hejhal's algorithm \cite{Hejhal1999}, which uses the Fourier expansion (\ref{TruncatedFourierExpansion}) and the automorphy condition (\ref{AutomorphyCondition}). In the present paper, we restrict ourselves to the modular group $\Gamma=\operatorname{PSL}(2,\mathbb{Z})$ which is generated by the translation $z\mapsto z+1$ and the inversion $z\mapsto -\frac{1}{z}$. There do not exist small eigenvalues $0<\lambda=r^2+\frac{1}{4}\le\frac{1}{4}$ for the modular group; see \cite{Roelcke1966}. Therefore, $r$ is real and the term $u_0(y)$ in the Fourier expansion of Maa{\ss} cusp forms (\ref{FourierExpansion}) vanishes. Due to the exponential decay of the K-Bessel function for large arguments (\ref{LargeArgumentK}), and the bound
\begin{align*} |a_n|\le d(n)n^{\frac{1}{4}} \end{align*}
for the coefficients, see \cite{Vigneras1983}, where $d(n)$ counts the number of divisors of $n$, the absolutely convergent Fourier expansion can be truncated anytime we bound $y$ from below. Given $\varepsilon>0$, $r$, and $y$, we determine the smallest $M=M(\varepsilon,r,y)$ such that the inequalities
\begin{align*} 2\pi M y\ge r \quad \text{and} \quad K_{\ii r}(2\pi M y)\le \varepsilon \max_{x}(K_{\ii r}(x)) \end{align*}
hold. Larger $y$ allow smaller $M$. In all the truncated terms, i.e. within
\begin{align*} [[\varepsilon]]=\sum_{n=M+1}^{\infty} a_n y^{\frac{1}{2}} K_{\ii r}(2\pi n y) \cs(2\pi n x), \end{align*}
the K-Bessel function decays exponentially in $n$, and already the K-Bessel function of the first truncated summand is smaller than $\varepsilon$ times most of the K-Bessel functions in the sum of (\ref{TruncatedFourierExpansion}). Thus, the error $[[\varepsilon]]$ does at most marginally exceed $\varepsilon$. The reason why $[[\varepsilon]]$ can exceed $\varepsilon$ somewhat is due to the possibility that the summands in (\ref{TruncatedFourierExpansion}) can cancel each other and that the first few coefficients $a_n$ in the truncated terms may occasionally be much bigger than in (\ref{TruncatedFourierExpansion}). \par
By a finite Fourier transform, the Fourier expansion (\ref{TruncatedFourierExpansion}) is solved for its coefficients
\begin{align} \label{FFT} a_m y^{\frac{1}{2}} K_{\ii r}(2\pi m y)=\frac{1}{2Q} \sum_{\substack{x\in\mathbb{X}}} f(x+\ii y) \cs(-2\pi m x) + [[\varepsilon]], \end{align}
where $\mathbb{X}$ is an equidistributed set of $Q$ numbers,
\begin{align*} \mathbb{X}=\{ \tfrac{\frac{1}{2}}{2Q},\tfrac{\frac{3}{2}}{2Q},\ldots,\tfrac{Q-\frac{3}{2}}{2Q},\tfrac{Q-\frac{1}{2}}{2Q} \}, \end{align*}
with $2Q>M+m$. \par
By automorphy we have
\begin{align*} f(z)=f(z^*), \end{align*}
where $z^*$ is the $\Gamma$-pullback of the point $z$ into the fundamental domain $\mathcal F$,
\begin{align*} z^*=\gamma z, \quad \gamma\in\Gamma, \quad z^*\in{\mathcal F}. \end{align*}
Any Maa{\ss} cusp form can thus be approximated by
\begin{align} \label{Pullback} f(x+\ii y)=f(x^*+\ii y^*)=\sum_{n=1}^{M_0} a_n {y^*}^{\frac{1}{2}} K_{\ii r}(2\pi n y^*) \cs(2\pi n x^*) + [[\varepsilon]], \end{align}
where $y^*$ is always larger or equal than the height $y_0$ of the lowest points in the fundamental domain $\mathcal F$,
\begin{align*} y_0=\min_{z\in{\mathcal F}}(y)=\frac{\sqrt{3}}{2}, \end{align*}
effectively allowing us to replace $M(\varepsilon,r,y)$ by $M_0=M(\varepsilon,r,y_0)$. \par
Choosing $y$ smaller than $y_0$, the $\Gamma$-pullback of any point into the fundamental domain $\mathcal F$ makes use at least once of the inversion $z\mapsto -\frac{1}{z}$, possibly together with the translation $z\mapsto z+1$. This is called implicit automorphy, since it guarantees the invariance $f(z)=f(-\frac{1}{z})$, whereas the condition $f(z)=f(z+1)$ is satisfied by the Fourier expansion. \par
Making use of the implicit automorphy by replacing $f(x+\ii y)$ in (\ref{FFT}) with the right-hand side of (\ref{Pullback}) yields
\begin{multline*} a_m y^{\frac{1}{2}} K_{\ii r}(2\pi m y) \\ =\frac{1}{2Q} \sum_{\substack{x\in\mathbb{X}}} \sum_{n=1}^{M_0} a_n {y^*}^{\frac{1}{2}} K_{\ii r}(2\pi n y^*) \cs(2\pi n x^*) \cs(-2\pi m x) + [[2\varepsilon]] \end{multline*}
for $1\le m\le M$, which is the central identity in the algorithm. With this identity, the coefficients $a_m$ can be determined for all $m$ so long as $y<y_0$ is chosen such that $K_{\ii r}(2\pi m y)$ does not become too small. \par
Taking $1\le m\le M_0$ and forgetting about the error $[[2\varepsilon]]$, the set of equations can be rewritten as
\begin{align} \label{Equations} \sum_{n=1}^{M_0} V_{mn}(r,y) a_n=0, \quad m\ge1, \end{align}
where the matrix $V=(V_{mn})$ is given by
\begin{multline*} V_{mn}(r,y)=y^{\frac{1}{2}} K_{\ii r}(2\pi m y) \delta_{mn} \\ - \frac{1}{2Q} \sum_{\substack{x\in\mathbb{X}}} {y^*}^{\frac{1}{2}} K_{\ii r}(2\pi n y^*) \cs(2\pi n x^*) \cs(-2\pi m x). \end{multline*}
Since $y$ can always be chosen such that $K_{\ii r}(2\pi m y)$ is not too small, the diagonal terms in the matrix $V$ do not vanish for large $m$ and the matrix is well conditioned. This makes the algorithm stable. \par
We are now looking for non-trivial solutions of (\ref{Equations}) with $1\le n\le M_0$ that simultaneously give the eigenvalues $r$ and the coefficients $a_n$. Trivial solutions are avoided by setting $a_1=1$, cf. \cite{Miyake1989}. \par
Since the eigenvalues $r$ are unknown, we discretize the $r$ axis and solve for each $r$ value on this grid,
\begin{align} \label{InhomogeneousEquations} \sum_{n=2}^{M_0} V_{mn}(r,y^{\#1}) a_n=-V_{m1}(r,y^{\#1}), \quad 1\le m\le M_0-1, \end{align}
where $y^{\#1}<y_0$ is chosen such that $K_{\ii r}(2\pi m y^{\#1})$ is not too small for $1\le m\le M_0-1$. A good value to try for $y^{\#1}$ is given by
\begin{align*} 2\pi M_0 y^{\#1}=r. \end{align*}
Hejhal \cite{Hejhal1999} solves (\ref{InhomogeneousEquations}) a second time with a different $y^{\#2}$ and checks whether the coefficients are independent of the choice of $y$. \par
Some words have to be said about what we mean by solving the inhomogeneous system (\ref{InhomogeneousEquations}), since it may happen that there is not always a solution unless $r$ is an eigenvalue. By solving a linear inhomogeneous system of equations
\begin{align*} Ax=y \end{align*}
we mean that we compute
\begin{align*} x=\tilde{A}^{-1}y \end{align*}
where $\tilde{A}^{-1}$ is determined such that $\tilde{A}^{-1}A$ is a diagonal matrix where as many diagonal elements as possible are equal to one.

\section{Some improvements} We restructure Hejhal's algorithm \cite{Hejhal1999} in the way it finds the eigenvalues. Instead of solving (\ref{InhomogeneousEquations}) a second time, we check whether the coefficients $a_n=a_n^{\#1}$ obtained actually solve (\ref{Equations}) by computing
\begin{align*} g_m=\sum_{n=1}^{M_0} V_{mn}(r,y^{\#2}) a_n^{\#1}, \quad 1\le m\le M_0, \end{align*}
where $y^{\#2}=\frac{9}{10}y^{\#1}$ is a good choice for an independent $y$ value. Only if all $g_m$ vanish simultaneously can the given $r$ be an eigenvalue and the computed $a_n$'s the Fourier coefficients of a Maa{\ss} cusp form. \par
The probability of finding an $r$ value such that all $g_m$ vanish simultaneously is zero because the discrete eigenvalues are of measure zero in the real numbers. Therefore, we make use of the intermediate value theorem. We let $r$ run through a grid of discretized $r$ values and look for {\it simultaneous changes of sign} in the $g_m$. \par
It is conjectured \cite{BogomolnyGeorgeotGiannoniSchmit1992,BolteSteilSteiner1992,Bolte1993,Sarnak1993,BogomolnyLeyvrazSchmit1996} that the eigenvalues of the Laplacian for even and odd cusp forms each possess a spacing distribution close to that of a Poisson random process. One therefore expects that small spacings will occur comparably often (due to level clustering). In order not to miss eigenvalues which lie close together, we have to make sure that at least one point of the $r$ grid lies between any two successive eigenvalues. On the other hand, we do not want to waste CPU time if there are large spacings. Therefore, we use an adaptive algorithm which tries to predict the next best $r$ value of the grid. It is based on the observation that the coefficients $a_n$ of two Maa{\ss} cusp forms of successive eigenvalues must differ. Assume that two eigenvalues lie close together and that the coefficients of the two Maa{\ss} cusp forms do not differ much. Numerically then both Maa{\ss} cusp forms would tend to be similar -- which contradicts the fact that different Maa{\ss} cusp forms are {\it orthogonal} to each other with respect to the Petersson scalar product
\begin{align*} \langle f_i,f_j \rangle = \int_{\mathcal F} \overline{f_i(z)} f_j(z) \frac{dx \, dy}{y^2} = 0, \quad \text{if } \lambda_i\not=\lambda_j. \end{align*}
Maa{\ss} cusp forms corresponding to different eigenvalues are orthogonal because the Laplacian is an essentially self-adjoint operator. Thus, if successive eigenvalues lie close together, the coefficients $a_n$ must change fast when varying $r$. In contrast, if successive eigenvalues are separated by large spacings numerically, it turns out that often the coefficients change only slowly upon varying $r$. Defining
\begin{align*} \tilde{a}_n=\frac{a_n}{\sqrt{\sum_{m=1}^{M_0} |a_m|^2}}, \quad 1\le n\le M_0, \end{align*}
our adaptive algorithm predicts the next $r$ value of the grid such that the change in the coefficients is
\begin{align} \label{Adaptive} \sum_{n=1}^{M_0} |\tilde{a}_n(r_{\text{old}})-\tilde{a}_n(r_{\text{new}})|^2\approx0.04. \end{align}
For this prediction, the last step in the $r$ grid together with the last change in the coefficients is used to extrapolate linearly the choice for the next $r$ value of the grid. \par
However the adaptive algorithm is not rigorous. Sometimes the prediction of the next $r$ value fails so that it is too close or too far away from the previous one. A small number of small steps does not bother us unless the step size tends to zero. But, if the step size is too large, such that the left-hand side of (\ref{Adaptive}) exceeds $0.16$, we reduce the step size and try again with a smaller $r$ value. \par
Compared to earlier algorithms, our adaptive one tends to miss significantly less eigenvalues per run. \par
We are searching for simultaneous sign changes in the quantities $g_m$. Once we have found such in at least half of all the $g_m$'s, we have found an interval $[r_{\text{old}},r_{\text{new}}]$ which contains an eigenvalue $r$ {\it with high probability}. The next step is to check whether this interval really contains an eigenvalue, and, if so, to find this eigenvalue by some interpolation or bisection. \par
In fact, we use a trisection which is based on a bisection together with a Newtonian interpolation. One first bisects the interval $[r_{\text{old}},r_{\text{new}}]$ and re-examines the sign changes. The interval with the most is then divided further by Newtonian interpolation, which ensures fast convergence. In the next step of the trisection, we again examine the sign changes and highlight that interval which contains the most simultaneous sign changes in the $g_m$'s. If there {\it is} an eigenvalue contained in the successive intervals of the trisection, the number of $g_m$'s that simultaneously change their sign increases from step to step in the iteration until the size of the interval approaches zero and the eigenvalue is found. In the opposite case, the number of $g_m$'s which simultaneously change their sign decreases from step to step in the iteration until one suspects that there is no eigenvalue contained in the interval $[r_{\text{old}},r_{\text{new}}]$. \par

\section{Results} After some preliminary tests of our algorithm, where we computed some eigenvalues of the odd symmetry class around $r\approx10000$, see table \ref{EigenvaluesOdd},
\begin{table} \caption{Eigenvalues of the Maa{\ss} cusp forms around $r\approx10000$ (odd symmetry).} \label{EigenvaluesOdd} \begin{align*} r& \\
10000&.00203541 \\
10000&.00469659 \\
10000&.00735313 \\
10000&.00773954 \\
10000&.00805085 \\
10000&.00947268 \\
10000&.01012235 \\
10000&.01102222 \\
10000&.01373844 \\
10000&.01460515 \\
10000&.01610617 \\
\end{align*} \end{table}
and two larger eigenvalues $r=20000.00164526$ (even) and $r=20000.00020183$ (odd), we decided to compute two Maa{\ss} cusp forms corresponding to eigenvalues $r\approx40000$, one for each symmetry class. For the size of the error in truncating the Fourier expansion, we chose $\varepsilon=10^{-7}$, which we took also for the accuracy of our K-Bessel function. For this cutoff, we had to take $7395$ Fourier coefficients into account of which $938$ have prime index. Finding the two eigenvalues together with their Maa{\ss} cusp forms took four weeks of CPU time for each on a $750$ MHz SUN UltraSPARC-III processor. $1.3$ GB of memory were needed. \par
Starting at $r=40000$ in the upwards direction, the first even eigenvalue was found at $r=40000.0000916$; the first odd one at $r=40000.0001644$. In tables \ref{CoefficientsEven}
\begin{table} \caption{The first $174$ Fourier coefficients of the Maa{\ss} cusp form corresponding to the eigenvalue $r=40000.0000916$ (even symmetry).} \label{CoefficientsEven} \begin{align*} a&_{1\ldots29} & a&_{30\ldots58} & a&_{59\ldots87} & a&_{88\ldots116} & a&_{117\ldots145} & a&_{146\ldots174} \\
1& & -0&.2674 & 0&.5078 & -0&.0963 & 0&.2399 & 1&.3396 \\
1&.2094 & 1&.4654 & -0&.1024 & 1&.1918 & 0&.6143 & 0&.0787 \\
-0&.1799 & -0&.8607 & -0&.5538 & -1&.4379 & 0&.2756 & 0&.6445 \\
0&.4629 & -0&.0267 & 1&.7726 & 0&.1861 & 0&.1435 & -1&.0355 \\
1&.2285 & -0&.4445 & 0&.7256 & -0&.6502 & -0&.9778 & -0&.1108 \\
-0&.2176 & -0&.9213 & 0&.2074 & -0&.2637 & -0&.6697 & -0&.6101 \\
-0&.7499 & -0&.4479 & -0&.3046 & 1&.1780 & -0&.1000 & 0&.7014 \\
-0&.6494 & 1&.3920 & -0&.0322 & -1&.3269 & 0&.6785 & 0&.3556 \\
-0&.9675 & -1&.3062 & -0&.6666 & 0&.1549 & -0&.6027 & -0&.1347 \\
1&.4858 & 0&.0446 & -0&.1701 & -0&.9765 & 0&.8777 & 1&.8005 \\
0&.1485 & -0&.7980 & 0&.2527 & -0&.5292 & 1&.6805 & 0&.0207 \\
-0&.0833 & 0&.5559 & -1&.1143 & -0&.1436 & 1&.1116 & -0&.5205 \\
-0&.2480 & 0&.1632 & 1&.2144 & 0&.2360 & -0&.2161 & 1&.2166 \\
-0&.9070 & 1&.2010 & 0&.6285 & -0&.3695 & -0&.3686 & 0&.0880 \\
-0&.2211 & 0&.0687 & 1&.1076 & 0&.0800 & 1&.5360 & -1&.0575 \\
-1&.2484 & -1&.1888 & 1&.6837 & -0&.5686 & -0&.0123 & 1&.0532 \\
-0&.3674 & -1&.6986 & -0&.0916 & 0&.1612 & 0&.8099 & 1&.0933 \\
-1&.1703 & 0&.9739 & -0&.5000 & 0&.1658 & -0&.8063 & -0&.8840 \\
-1&.0799 & 0&.2246 & -0&.1113 & -0&.5918 & 0&.4349 & 0&.2573 \\
0&.5687 & -0&.4376 & 0&.0540 & 0&.7798 & 0&.2387 & -0&.0329 \\
0&.1349 & 0&.6161 & 1&.0058 & 0&.1639 & 1&.3926 & -2&.0562 \\
0&.1795 & 0&.0661 & -1&.5339 & 0&.8569 & 0&.3056 & -0&.3518 \\
-1&.4044 & -0&.1149 & 0&.9039 & 0&.2205 & -1&.0796 & -0&.0877 \\
0&.1169 & -0&.4892 & 0&.6724 & -0&.2505 & -0&.4266 & -0&.9383 \\
0&.5094 & 0&.4282 & -1&.6999 & 0&.9363 & -0&.1751 & -0&.5461 \\
-0&.3000 & 0&.1824 & 0&.0624 & 1&.7346 & 1&.4690 & 1&.0451 \\
0&.3540 & 0&.4871 & -0&.4514 & 0&.2351 & -0&.0367 & 0&.5561 \\
-0&.3473 & 0&.1943 & 1&.4526 & -1&.7254 & 1&.2081 & -0&.4913 \\
0&.2914 & 0&.3524 & -0&.0524 & 0&.1349 & 0&.3580 & -0&.0634 \\
\end{align*} \end{table}
and \ref{CoefficientsOdd},
\begin{table} \caption{The first $174$ Fourier coefficients of the Maa{\ss} cusp form corresponding to the eigenvalue $r=40000.0001644$ (odd symmetry).} \label{CoefficientsOdd} \begin{align*} a&_{1\ldots29} & a&_{30\ldots58} & a&_{59\ldots87} & a&_{88\ldots116} & a&_{117\ldots145} & a&_{146\ldots174} \\
1& & -0&.1394 & 0&.9151 & 1&.2840 & -0&.1093 & 0&.6031 \\
-0&.5454 & 0&.7975 & -0&.1794 & -1&.6376 & -0&.4991 & -0&.7992 \\
0&.8637 & -1&.0355 & -1&.6485 & 0&.0410 & 0&.4208 & 0&.6505 \\
-0&.7026 & 1&.1944 & -0&.4350 & 0&.1176 & 0&.2371 & 0&.5716 \\
0&.2957 & -0&.8396 & -0&.0695 & 0&.1591 & 0&.9124 & 0&.4298 \\
-0&.4711 & 0&.0809 & 0&.3686 & 0&.6888 & 0&.8991 & -1&.6099 \\
0&.2733 & 0&.1785 & 0&.1271 & 0&.0114 & -1&.5822 & 1&.4495 \\
0&.9285 & -0&.9259 & -0&.6514 & 0&.4616 & -0&.5604 & -0&.3911 \\
-0&.2540 & -0&.8514 & 1&.1840 & -0&.8943 & -0&.5655 & -0&.2062 \\
-0&.1612 & 0&.3713 & -1&.0815 & 1&.8048 & 0&.0378 & 0&.2358 \\
1&.3829 & 0&.2746 & -0&.1956 & 0&.5047 & -1&.0520 & -0&.2609 \\
-0&.6068 & -1&.8319 & -0&.0441 & -0&.3512 & 0&.8345 & -0&.6977 \\
0&.4299 & -0&.1287 & 1&.8049 & 0&.6412 & -1&.1665 & -0&.3762 \\
-0&.1491 & -1&.3506 & -0&.2359 & -0&.0862 & -0&.0694 & 1&.2012 \\
0&.2554 & -0&.9716 & -1&.1057 & -0&.7252 & -0&.6646 & -0&.3062 \\
0&.1962 & -0&.0751 & 0&.5050 & 0&.4610 & -0&.8391 & -0&.0619 \\
1&.5395 & 0&.1235 & -0&.7882 & 0&.3993 & 0&.4267 & 0&.3716 \\
0&.1386 & -0&.0211 & -1&.0967 & 0&.0698 & -0&.6457 & -1&.0045 \\
1&.5611 & 0&.1694 & 0&.3780 & -0&.7585 & -0&.3202 & 1&.2870 \\
-0&.2077 & -0&.9253 & -0&.2025 & -1&.8158 & 1&.4295 & 0&.3531 \\
0&.2361 & 0&.4977 & 0&.6896 & 0&.7609 & 0&.5571 & -0&.3239 \\
-0&.7542 & 1&.3296 & 0&.0580 & -0&.2113 & 0&.1068 & -0&.1377 \\
-0&.2266 & -0&.3019 & -0&.6815 & -0&.2230 & 0&.2784 & 0&.2192 \\
0&.8019 & 1&.3908 & 0&.9990 & -0&.7997 & -0&.0568 & -0&.8151 \\
-0&.9126 & 0&.5907 & 0&.5937 & 0&.0536 & -0&.0182 & -0&.2483 \\
-0&.2344 & 0&.4089 & -0&.1658 & 1&.7004 & -0&.9844 & -0&.3966 \\
-1&.0831 & 0&.2538 & 0&.4552 & -0&.7353 & 0&.5945 & 0&.9489 \\
-0&.1921 & 1&.3483 & 0&.7366 & -0&.0670 & -0&.0499 & -0&.3371 \\
-0&.3333 & 0&.1817 & -0&.2879 & 0&.2341 & -0&.0985 & 0&.1570 \\
\end{align*} \end{table}
we list the first few Fourier coefficients of these two forms. We checked the accuracy of our results with the aid of the multiplicative relations (\ref{MultiplicativeRelations}). The left-hand side of the multiplicative relations coincides with the right-hand side up to a discrepancy of size $10^{-3}$. This means that the coefficients are only accurate to {\it three} digits. This is much worse than the initially intended accuracy $\varepsilon=10^{-7}$. The reason for this loss of accuracy is that we have computed the eigenvalues only up to an accuracy of twelve digits. Minimal deviations of the eigenvalue $r$ lead to big changes in the coefficients $a_n$. But we cannot compute the eigenvalue much more accurately without increasing the accuracy of our K-Bessel routine and taking more coefficients in the Fourier expansion into account. \par
A different check of the accuracy of the results can be done by computing the coefficients $a_n$ a second time, with $y^{\#1}$ replaced by $y^{\#2}$. But, with this check, one has to be careful because the coefficients may vary less than the size of their actual error. We did this check and found that the coefficients differ in the sixth digit when $y^{\#1}$ is replaced by $y^{\#2}$. \par
\begin{figure} \psfrag{drho/du}{$\frac{d\nu}{du}$} \psfrag{u}{$u$} \psfrag{rho}{$\nu$} \includegraphics{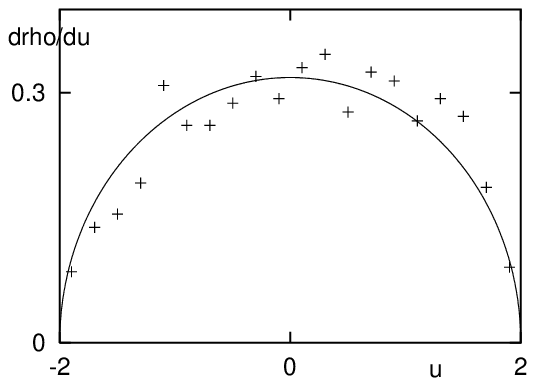} \hfill \includegraphics{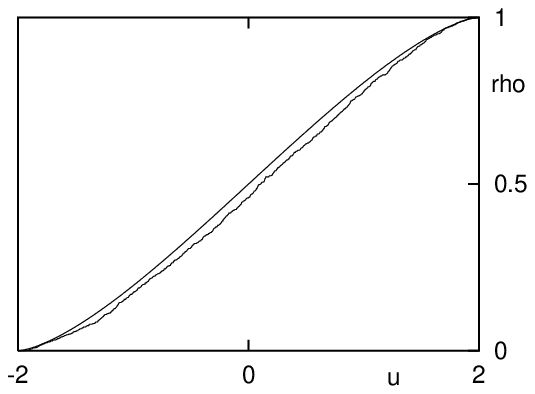} \caption{Statistics of the first $938$ prime Fourier coefficients of the Maa{\ss} cusp form corresponding to the eigenvalue $r=40000.0000916$. In the left figure the distribution of the prime coefficients is rendered with points. The solid line is the conjectured semicircle. In the right figure the solid line is the integrated distribution of the prime coefficients, and the dashed line is the integrated semicircle.} \label{SatoTateEven} \end{figure}
\begin{figure} \psfrag{drho/du}{$\frac{d\nu}{du}$} \psfrag{u}{$u$} \psfrag{rho}{$\nu$} \includegraphics{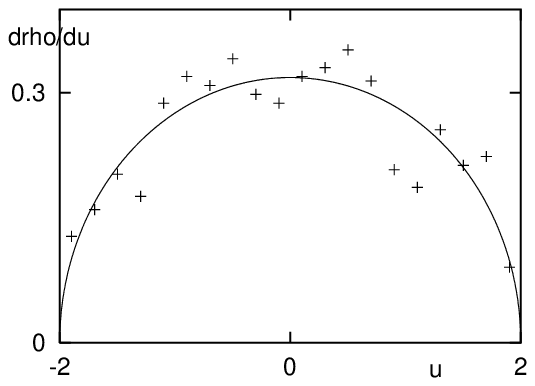} \hfill \includegraphics{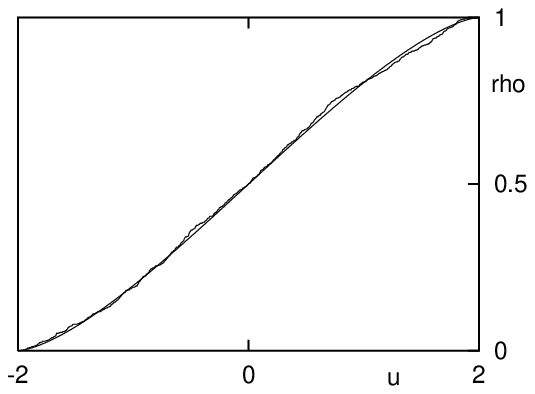} \caption{Statistics of the first $938$ prime Fourier coefficients of the Maa{\ss} cusp form corresponding to the eigenvalue $r=40000.0001644$. In the left figure the distribution of the prime coefficients is rendered with points. The solid line is the conjectured semicircle. In the right figure the solid line is the integrated distribution of the prime coefficients, and the dashed line is the integrated semicircle.} \label{SatoTateOdd} \end{figure}

All coefficients which we have computed satisfy the Ramanujan-Petersson conjecture
\begin{align*} |a_p|\le2 \quad \text{for all primes $p$}. \end{align*}
If the Sato-Tate conjecture is true, the prime coefficients $a_p$ of each Maa{\ss} cusp form are distributed according to the semicircle law
\begin{align*} d\nu(u)=\begin{cases} \frac{1}{2\pi}\sqrt{4-u^2}\,du & \text{if $|u|<2$}, \\ 0 & \text{otherwise}. \end{cases} \end{align*}
This means that
\begin{align*} \lim_{N\to\infty} \frac{\frac{1}{\#\{p \text{ prime};\ p\le N\}} \sum_{\substack{p\le N\\p \text{ prime}}} \chi_{[a,b]}(a_p)}{\int_a^b\,d\nu(u)}=1 \end{align*}
holds for any $-\infty<a<b<\infty$, where $\chi_{[a,b]}(u)$ is the indicator function of the interval $[a,b]$. The prime coefficients which we have computed match the Sato-Tate conjecture moderately well; see figures \ref{SatoTateEven}
and \ref{SatoTateOdd}.
One expects of course that the Sato-Tate conjecture {\it is} true and that the plots rapidly improve once one takes more coefficients (with $p>M_0$) into account; cf. \cite{HejhalArno1993,Steil1994}.

\section{Value distribution} 
\begin{figure} \psfrag{drho/du}{$\frac{d\rho}{du}$} \psfrag{u}{$u$} \psfrag{rho}{$\rho$} \includegraphics{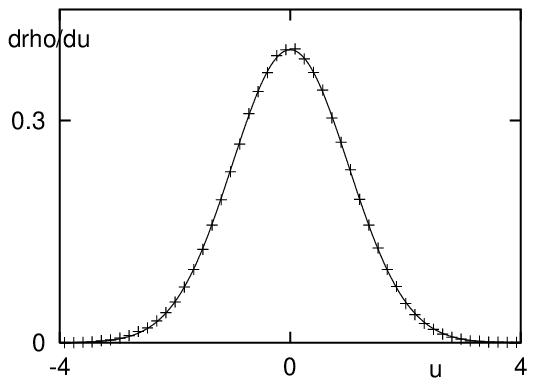} \hfill \includegraphics{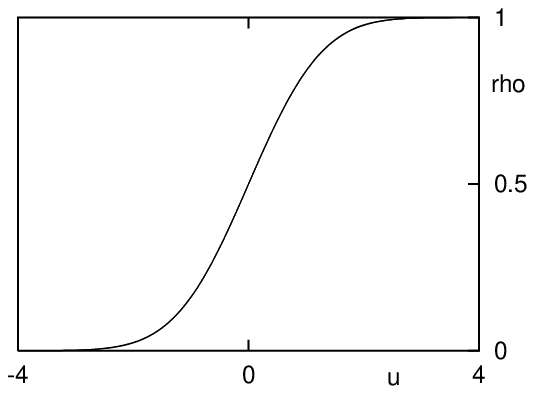} \caption{In the left figure the value distribution of the Maa{\ss} cusp form corresponding to the eigenvalue $r=40000.0000916$ inside the region $F$ is rendered with points. The solid line is the conjectured Gaussian. In the right figure the solid line is the integrated value distribution of the Maa{\ss} cusp form which is indistinguishable from the integrated Gaussian.} \label{BerryEven} \end{figure}

\begin{figure} \psfrag{drho/du}{$\frac{d\rho}{du}$} \psfrag{u}{$u$} \psfrag{rho}{$\rho$} \includegraphics{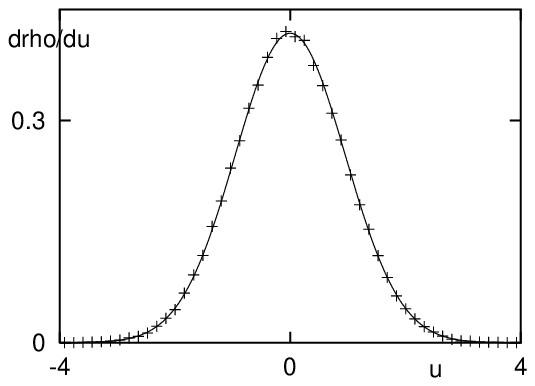} \hfill \includegraphics{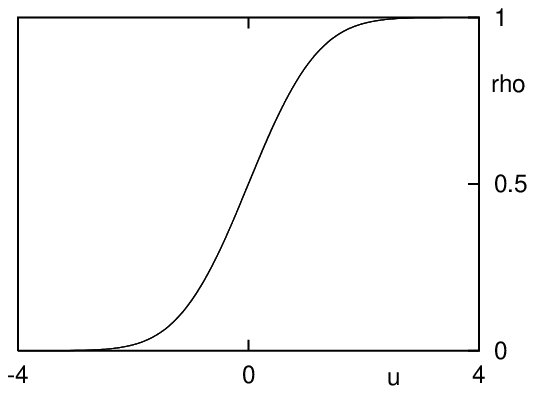} \caption{In the left figure the value distribution of the Maa{\ss} cusp form corresponding to the eigenvalue $r=40000.0001644$ inside the region $F$ is rendered with points. The solid line is the conjectured Gaussian. In the right figure the solid line is the integrated value distribution of the Maa{\ss} cusp form which is indistinguishable from the integrated Gaussian.} \label{BerryOdd} \end{figure}

It is believed that Maa{\ss} cusp forms behave pretty much like random waves. In particular, in the limit of large eigenvalues, $\lambda=r^2+\frac{1}{4}\to\infty$, a conjecture of Berry \cite{Berry1977} predicts that each Maa{\ss} cusp form has a Gaussian value distribution,
\begin{align*} d\rho(u)=\frac{1}{\sqrt{2\pi}\sigma}\e^{-\frac{u^2}{2\sigma^2}}\,du, \end{align*}
inside any compact regular subregion $F$ of $\mathcal F$. This means that
\begin{align*} \lim_{\lambda\to\infty} \frac{\frac{1}{\operatorname{area}(F)} \int_F \chi_{[a,b]}(f(z))\,d\mu}{\int_a^b\,d\rho(u)}=1 \end{align*}
holds with variance
\begin{align*} \sigma^2=\frac{1}{\operatorname{area}(F)} \int_F |f(z)|^2\,d\mu \end{align*}
for any $-\infty<a<b<\infty$. Figures \ref{BerryEven}
and \ref{BerryOdd}
show the value distribution of the Maa{\ss} cusp forms corresponding to the eigenvalues $r=40000.0000916$ resp. $r=40000.0001644$ inside a small subregion
\begin{align*} F=\{z=x+\ii y; \quad -0.3\le x\le -0.29215, \quad 1.1\le y\le 1.10785\}. \end{align*}
(See \cite{Hejhal1999} for some analogous plots with smaller $r$.) \par
\begin{figure} \includegraphics[width=4.99in,angle=-90]{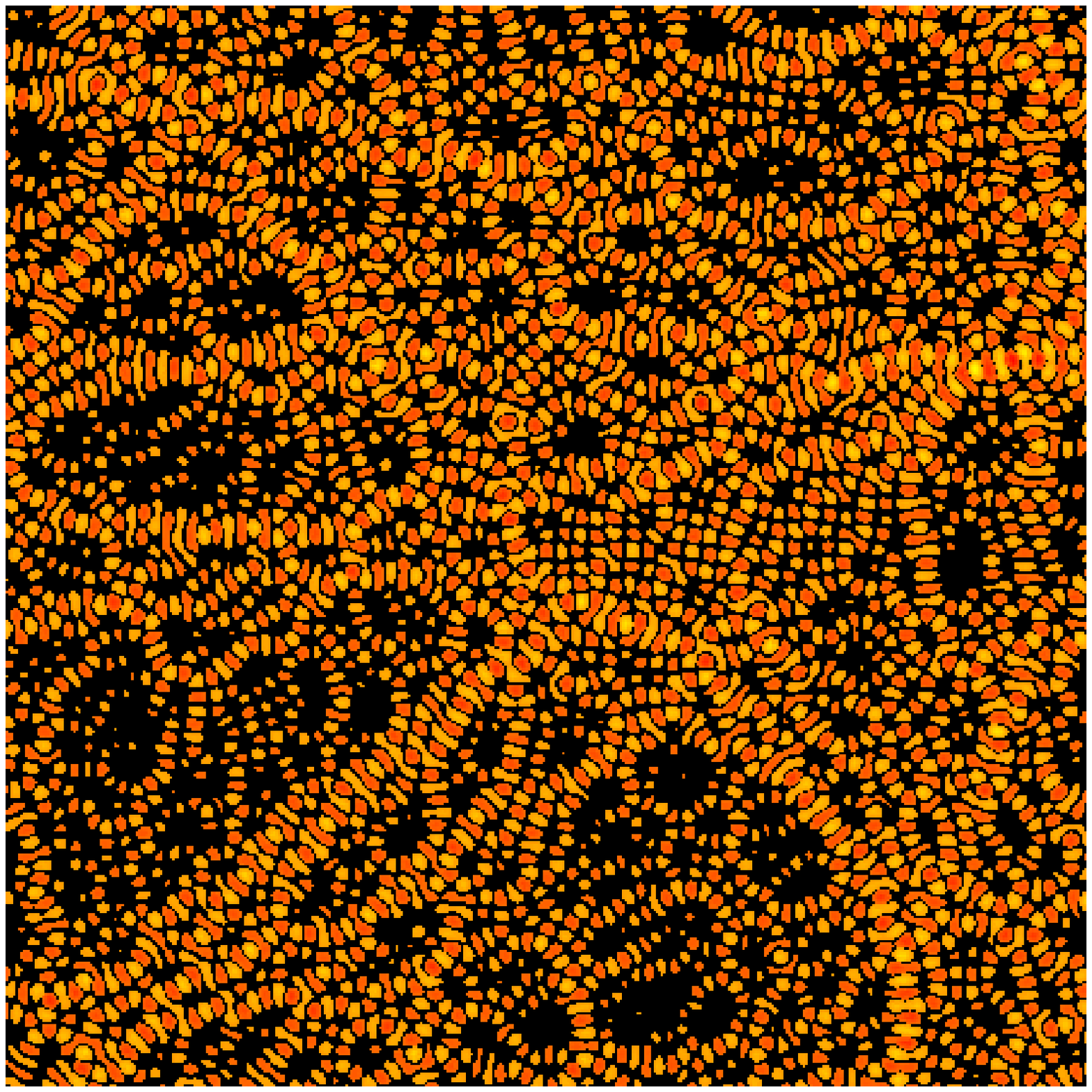} \caption{A plot of the Maa{\ss} cusp form corresponding to the eigenvalue $r=40000.0000916$ inside the region $F$.} \label{PlotEven} \end{figure}
\begin{figure} \includegraphics[width=4.99in,angle=-90]{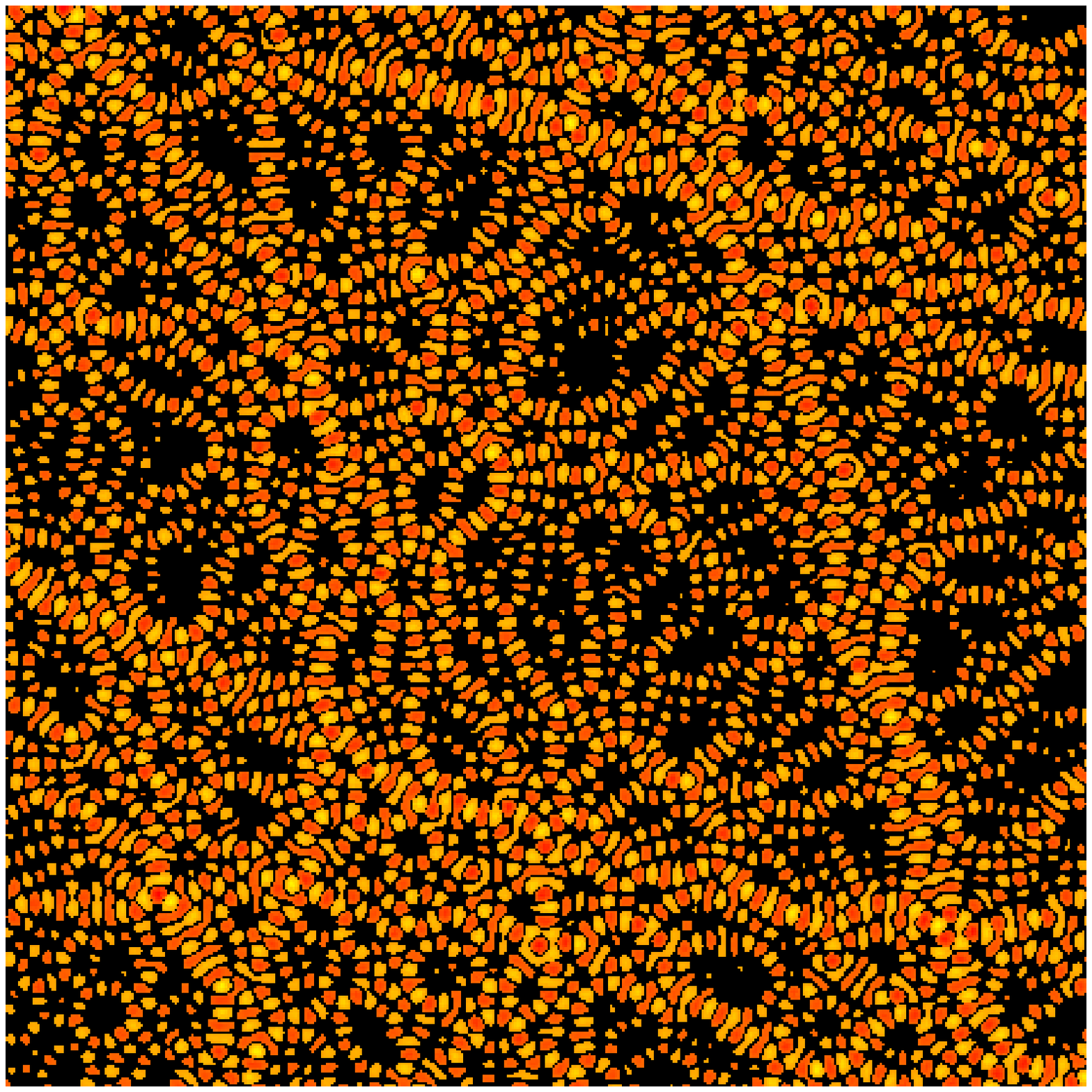} \caption{A plot of the Maa{\ss} cusp form corresponding to the eigenvalue $r=40000.0001644$ inside the region $F$.} \label{PlotOdd} \end{figure}
\begin{figure} \psfrag{F}{$F$} \includegraphics{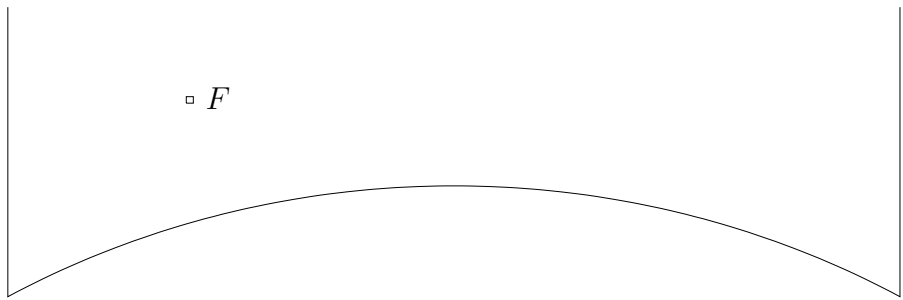} \caption{The small subregion $F$ inside the fundamental domain $\mathcal F$.} \label{RegionF} \end{figure}

Our numerical data agree well with Berry's conjecture, providing additional numerical evidence that the conjecture does hold. Plots of the two Maa{\ss} cusp forms inside the region $F$ are given in figures \ref{PlotEven}
and \ref{PlotOdd}.
Figure \ref{RegionF}
shows the small region $F$ inside the fundamental domain $\mathcal F$.

\appendix
\section{The K-Bessel function} \label{KBessel} 
\begin{figure} \psfrag{K}{$\e^{\frac{\pi r}{2}}K_{\ii r}$} \psfrag{x}{$x$} \includegraphics{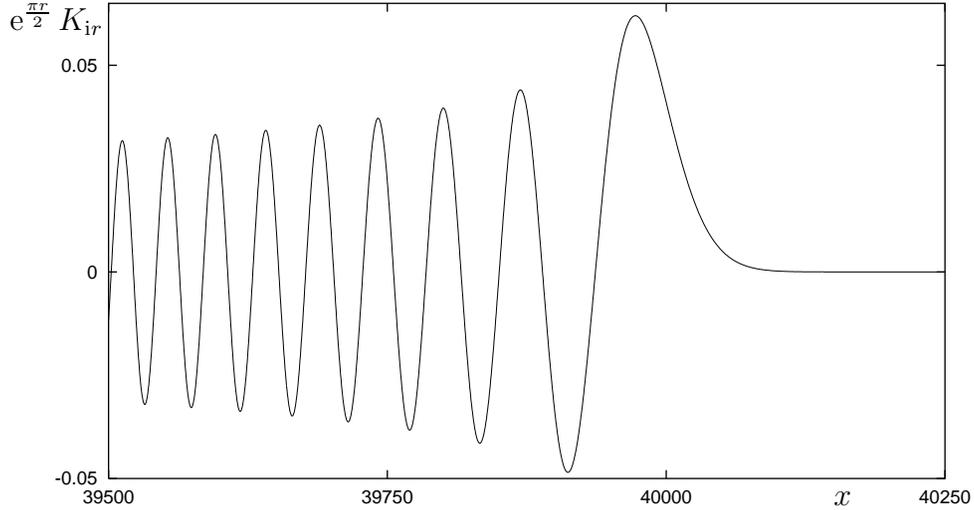} \caption{$\e^{\frac{\pi r}{2}} K_{\ii r}(x)$ for fixed $r=40000$.} \label{GraphK} \end{figure}

The K-Bessel function is defined by
\begin{align*} K_{\ii r}(x)=\int_{0}^{\infty} \e^{-x\cosh t} \cos(rt) \, dt, \quad \Re x>0, \quad r\in\mathbb{C}, \end{align*}
see Watson \cite{Watson1944}, and is real for real arguments $x$ and real or imaginary order $\ii r$. It satisfies the modified Bessel differential equation
\begin{align*} x^2 u''(x) + x u'(x) - (x^2 - r^2) u(x) = 0, \end{align*}
and decays exponentially for large arguments
\begin{align} \label{LargeArgumentK} K_{\ii r}(x)\sim\sqrt{\frac{\pi}{2x}} \e^{-x} \quad \text{for } x\to\infty. \end{align}
A second linearly independent solution of the differential equation is the I-Bessel function
\begin{align*} I_{\ii r}(x)=(\frac{x}{2})^{\ii r} \sum_{k=0}^{\infty} \frac{(\frac{x}{2})^{2k}}{k!\Gamma(\ii r+k+1)}, \end{align*}
which grows exponentially for large arguments
\begin{align*} I_{\ii r}(x)\sim\sqrt{\frac{1}{2\pi x}} \e^{x} \quad \text{for } x\to\infty. \end{align*}
The amplitude of the K-Bessel function gets exponentially small if $r$ increases. This can be compensated for by multiplication with the factor $\e^{\frac{\pi r}{2}}$, see figure \ref{GraphK}.
To compute the K-Bessel function numerically we use asymptotic expansions for large imaginary order, $r\to\infty$. The most powerful among them is the uniform asymptotic expansion
\begin{multline*} \e^{\frac{\pi r}{2}}K_{\ii r}(x)\sim 2^{\frac{1}{2}}\pi\Bigl(\frac{-\xi}{r^2-x^2}\Bigr)^{\frac{1}{4}}\biggl(Ai(\xi)\sum_{k=0}^{\infty}\frac{A_k(r^{-\frac{2}{3}}\xi)}{r^{2k}} \\ +Ai'(\xi)\sum_{k=0}^{\infty}\frac{B_k(r^{-\frac{2}{3}}\xi)}{r^{2k+\frac{4}{3}}}\biggr) \end{multline*}
where $Ai(x)$ and $Ai'(x)$ denote the Airy function and its derivative respectively. $\xi$ is defined by
\begin{align*} \beta=\frac{x}{r}, \quad \gamma=\frac{1}{\sqrt{1-\beta^2}}, \quad -\frac{2}{3r}(-\xi)^{\frac{3}{2}}=\gamma^{-1}-\sech^{-1}\beta, \end{align*}
and the functions $A_k(x)$ and $B_k(x)$ are given by
\begin{align*} \frac{A_k(r^{-\frac{2}{3}}\xi)}{r^{2k}}&=(-1)^{k}\sum_{s=0}^{2k}\frac{1+6s}{1-6s}\lambda_s(-\xi)^{-\frac{3}{2}s}\frac{u_{2k-s}(\gamma)}{r^{2k-s}}, \\ \frac{B_k(r^{-\frac{2}{3}}\xi)}{r^{2k+\frac{4}{3}}}&=(-1)^{k+1}\sum_{s=0}^{2k+1}\lambda_s(-\xi)^{-\frac{3}{2}s-\frac{1}{2}}\frac{u_{2k+1-s}(\gamma)}{r^{2k+1-s}}, \end{align*}
where
\begin{align*} \lambda_0=1, \quad \lambda_1=\frac{5}{48}, \quad \lambda_s=\frac{(6s-5)(6s-1)}{48s}\lambda_{s-1}, \quad s\ge2, \end{align*}
and $u_k(t)$ are polynomials satisfying the recursion
\begin{align*} &u_0(t)=1, \\ &u_{k+1}(t)=\frac{1}{2}t^2(1-t^2)u_k'(t)+\frac{1}{8}\int_{0}^{t}(1-5t^2)u_k(t)\, dt, \quad k\ge0, \end{align*}
see e.g. \cite[eq. (2)]{Balogh1966}, \cite[eqs. (18) (19) (20)]{Balogh1967}, \cite[appendix]{CsordasGrahamSzepfalusy1991}, \cite[section 5]{GilSeguraTemme2002}, or compare with \cite[eq. (4.24)]{Olver1954}, \cite[eq. (6.6)]{ChesterFriedmanUrsell1956}, \cite[eqs. (9.3.10) (9.3.35) (9.3.40) (9.3.41)]{AbramowitzStegun1965}. All terms are real if $x<r$, and using
\begin{align*} \frac{2}{3r}\xi^{\frac{3}{2}}=(-\ii\gamma)^{-1}-\sec^{-1}\beta, \quad \tilde{u}_k(-\ii\gamma)=(-\ii)^{k}u_k(\gamma), \end{align*}
all terms are real if $x>r$ with positive $-\ii\gamma$. Numerically, the uniform asymptotic expansion breaks down if $x$ comes close to $r$. But since it is analytic one can expand it around the transitional point $x=r$ and obtains
\begin{multline*} \e^{\frac{\pi r}{2}}K_{\ii r}(r-tr^{\frac{1}{3}})\sim\pi\biggl(\bigl(\frac{2}{r}\bigr)^{\frac{1}{3}}Ai(-2^{\frac{1}{3}}t)\sum_{k=0}^{\infty}(-1)^{k}\frac{\tilde{A}_k(t)}{r^{\frac{2k}{3}}} \\ +\bigl(\frac{4}{r}\bigr)^{\frac{1}{3}}Ai'(-2^{\frac{1}{3}}t)\sum_{k=0}^{\infty}(-1)^{k}\frac{\tilde{B}_k(t)}{r^{\frac{2k}{3}}}\biggr), \quad \text{$t$ small}, \end{multline*}
where the polynomials $\tilde{A}_k(t)$ and $\tilde{B}_k(t)$ are given in \cite[eq. (2.42)]{Olver1952}. Another useful asymptotic expansion in the transitional region is the Nicholson series \cite[p. 145]{MagnusOberhettingerSoni1966}, \cite[eq. (8)]{Balogh1967}
\begin{align*} \e^{\frac{\pi r}{2}}K_{\ii(x-t x^{\frac{1}{3}})}(x)\sim\pi\bigl(\frac{2}{x}\bigr)^{\frac{1}{3}}P(x,t)Ai\bigl(Q(x,t)\bigr), \quad \text{$t$ small}, \end{align*}
where the functions $P(x,t)$ and $Q(x,t)$ are defined by
\begin{align*} P(x,t)=\sum_{k=0}^{\infty}\bigl(\frac{2}{x}\bigr)^{\frac{2k}{3}}p_k(2^{\frac{1}{3}}t), \quad Q(x,t)=\sum_{k=0}^{\infty}\bigl(\frac{2}{x}\bigr)^{\frac{2k}{3}}q_k(2^{\frac{1}{3}}t), \end{align*}
and the polynomials $p_k(t)$ and $q_k(t)$ are given in \cite[p. 290]{Schoebe1954}. Substituting asymptotic expansions of the Airy function in the uniform asymptotic expansion of the K-Bessel function leads to the Hankel series
\begin{multline*} \e^{\frac{\pi r}{2}}K_{\ii r}(x)\sim\sqrt{2\pi\gamma r}\biggl(\sin\Bigl(\frac{2}{3}(-\xi)^{\frac{3}{2}}+\frac{\pi}{4}\Bigr)\sum_{k=0}^{\infty}\frac{(-1)^{k}}{r^{2k}}u_{2k}(\gamma) \\ +\cos\Bigl(\frac{2}{3}(-\xi)^{\frac{3}{2}}+\frac{\pi}{4}\Bigr)\sum_{k=0}^{\infty}\frac{(-1)^{k}}{r^{2k+1}}u_{2k+1}(\gamma)\biggr), \quad x\ll r, \end{multline*}
and the Debye series
\begin{align*} \e^{\frac{\pi r}{2}}K_{\ii r}(x)\sim\sqrt{2\pi(-\ii\gamma)r}\biggl(\frac{1}{2}\exp\Bigl(-\frac{2}{3}\xi^{\frac{2}{3}}\Bigr)\sum_{k=0}^{\infty}\frac{1}{r^{k}}\tilde{u}_{k}(-\ii\gamma)\biggr), \quad x\gg r, \end{align*}
see e.g. \cite[eqs. (2.14) (2.19)]{Olver1954}, \cite[eqs. (9.7.8) (9.3.10)]{AbramowitzStegun1965}, \cite[eqs. (3) (5)]{Balogh1967}, \cite[eqs. (A9) (A10)]{CsordasGrahamSzepfalusy1991}. Numerically, we tested all the given asymptotic expansions against each other to find out their range of applicability and their accuracy. We found by using the first five summands in the Hankel, the Debye and the Nicholson series, respectively, that the K-Bessel function can be approximated with an accuracy of {\it at least $10$} digits for $r\approx40000$ and all $x>0$.


\begin{thebibliography}{BGGS92}

\bibitem[AS65]{AbramowitzStegun1965}
M.~Abramowitz and I.~A. Stegun, \emph{Handbook of Mathematical Functions},
  Dover, 1965.

\bibitem[ASD71]{AtkinSwinnerton1971}
A.~O.~L. Atkin and H.~P.~F. Swinnerton-Dyer, \emph{Modular forms on
  noncongruence subgroups}, Combinatorics (Proc. Sympos. Pure Math., Vol. XIX,
  Univ. California, Los Angeles, Calif., 1968), Amer. Math. Soc. (1971),
  1--25.

\bibitem[Ave03]{Avelin2003}
H.~Avelin, \emph{On the deformation of cusp forms (Licentiate Thesis)}, UUDM
  report 2003:8 (Uppsala 2003).

\bibitem[Bal66]{Balogh1966}
C.~B. Balogh, \emph{Uniform asymptotic expansions of the modified Bessel
  function of the third kind of large imaginary order}, Bull. Amer. Math. Soc.
  \textbf{72} (1966), 40--43.

\bibitem[Bal67]{Balogh1967}
C.~B. Balogh, \emph{Asymptotic expansions of the modified Bessel function
  of the third kind of imaginary order}, SIAM J. Appl. Math. \textbf{15}
  (1967), no.~5, 1315--1323.

\bibitem[Ber77]{Berry1977}
M.~V. Berry, \emph{Regular and irregular semiclassical wavefunctions}, J.
  Phys. A \textbf{10} (1977), 2083--2091.

\bibitem[BGGS92]{BogomolnyGeorgeotGiannoniSchmit1992}
E.~Bogomolny, B.~Georgeot, M.-J. Giannoni, and C.~Schmit, \emph{Chaotic
  billiards generated by arithmetic groups}, Phys. Rev. Lett.
  \textbf{69} (1992), 1477--1480.

\bibitem[BLS96]{BogomolnyLeyvrazSchmit1996}
E.~Bogomolny, F.~Leyvraz, and C.~Schmit, \emph{Distribution of eigenvalues
  for the modular group}, Comm. Math. Phys. \textbf{176} (1996), no.~3,
  577-617.

\bibitem[Bol93]{Bolte1993}
J.~Bolte, \emph{Some studies on arithmetical chaos in classical and quantum
  mechanics}, Int. J. Mod. Phys. B \textbf{7} (1993), 4451--4553.

\bibitem[BSS92]{BolteSteilSteiner1992}
J.~Bolte, G.~Steil, and F.~Steiner, \emph{Arithmetical chaos and violation of
  universality in energy level statistics}, Phys. Rev. Lett.
  \textbf{69} (1992), 2188--2191.

\bibitem[Car71]{Cartier1971}
P.~Cartier, \emph{Some numerical computations relating to automorphic
  functions}, Computers in Number Theory (A.~O.~L. Atkin and B.~J. Birch,
  eds.), Academic Press, 1971, pp.~37--48.

\bibitem[Car78]{Cartier1978}
P.~Cartier, \emph{Analyse num\'{e}rique d'un probl\`{e}me de valeurs
  propres a haute pr\'{e}cision [application aux fonctions automorphes]},
  preprint, IHES (1978).

\bibitem[CFU57]{ChesterFriedmanUrsell1956}
C.~Chester, B.~Friedman, and F.~Ursell, \emph{An extension of the method of
  steepest descents}, Proc. Camb. Phil. Soc. \textbf{53} (1957), 599--611.

\bibitem[CGS91]{CsordasGrahamSzepfalusy1991}
A.~Csord\'{a}s, R.~Graham, and P.~Sz\'{e}pfalusy, \emph{Level statistics of
  a noncompact cosmological billiard}, Phys. Rev. A \textbf{44} (1991),
  1491--1499.

\bibitem[GS82]{GolovcanskiiSmotrov1982}
V.~V. Golov\v{c}anski\v{\i} and M.~N. Smotrov, \emph{The first few
  eigenvalues of the Laplacian on the fundamental domain of the modular
  group}, preprint, Far Eastern Scientific Center, Academy of Science USSR,
  Wladiwostok (1982) (Russian).

\bibitem[GS84]{GolovcanskiiSmotrov1984}
V.~V. Golov\v{c}anski\v{\i} and M.~N. Smotrov, \emph{Calculation of first
  Fourier coefficients of eigenfunctions of the Laplace operator on the
  fundamental domain of a modular group}, Numerical methods in algebra and
  analysis, Akad. Nauk SSSR, Dal. nevostochn. Nauchn. Tsentr, Vladivostok
  \textbf{85} (1984), 15--19 (Russian).

\bibitem[GST02]{GilSeguraTemme2002}
A.~Gil, J.~Segura, and N.~M. Temme, \emph{Computation of the modified
  Bessel function of the third kind of imaginary orders: Uniform
  Airy-type asymptotic expansion}, CWI report MAS-R0205 (2002).

\bibitem[HA93]{HejhalArno1993}
D.~A. Hejhal and S.~Arno, \emph{On Fourier coefficients of Maass
  waveforms for $\operatorname{PSL}(2,\mathbb{Z})$}, Math. Comp. \textbf{61}
  (1993), 245--267.

\bibitem[Haa77]{Haas1977}
H.~Haas, \emph{Numerische Berechnung der Eigenwerte der
  Differential\-glei\-chung $-{\Delta} u=\lambda y^{-2} u$ f\"{u}r ein
  unendliches Gebiet im $\mathbb{R}^2$}, 1977, Diplomarbeit,
  Universit\"{a}t Heidelberg, Institut f\"{u}r Angewandte Mathematik.

\bibitem[HB82]{HejhalBerg1982}
D.~A. Hejhal and B.~Berg, \emph{Some new results concerning eigenvalues of
  the non-Euclidean Laplacian for $\operatorname{PSL}(2,\mathbb{Z})$},
  Tech. report 82-172, University of Minnesota, 1982.

\bibitem[Hej81]{Hejhal1981}
D.~A. Hejhal, \emph{Some observations concerning eigenvalues of the
  Laplacian and Dirichlet L-series}, Resent Progress in Analytic
  Number Theory (H.~Halberstam and C.~Hooley, eds.), Academic Press, 1981,
  pp.~95--110.

\bibitem[Hej83]{Hejhal1983}
D.~A. Hejhal, \emph{The Selberg trace formula for
  $\operatorname{PSL}(2,\mathbb{R})$}, Lecture Notes in Math. 1001,
  Springer, 1983.

\bibitem[Hej91]{Hejhal1991}
D.~A. Hejhal, \emph{Eigenvalues for the Laplacian for
  $\operatorname{PSL}(2,\mathbb{Z})$: some new results and computational
  techniques}, International Symposium in Memory of Hua Loo-Keng
  (S.~Gong, Q.~K. Lu, Y.~Wang, and L.~Yang, eds.), Springer, 1991, pp.~59--102.

\bibitem[Hej92a]{Hejhal1992a}
D.~A. Hejhal, \emph{Eigenvalues of the Laplacian for Hecke triangle
  groups}, Mem. Amer. Math. Soc. \textbf{469} (1992).

\bibitem[Hej92b]{Hejhal1992b}
D.~A. Hejhal, \emph{On eigenvalues of the Laplacian for Hecke triangle
  groups}, Zeta Functions in Geometry (N.~Kurokawa and T.~Sunada, eds.),
  vol.~21, Adv. Stud. Pure Math., 1992, pp.~359--408.

\bibitem[Hej99]{Hejhal1999}
D.~A. Hejhal, \emph{On eigenfunctions of the Laplacian for Hecke triangle
  groups}, Emerging applications of number theory (D.~A. Hejhal, J.~Friedman,
  M.~C. Gutzwiller, and A.~M. Odlyzko, eds.), IMA Series No. 109, Springer,
  1999, pp.~291--315.

\bibitem[HR92]{HejhalRackner1992}
D.~A. Hejhal and B.~Rackner, \emph{On the topography of Maass waveforms
  for $\operatorname{PSL}(2,\mathbb{Z})$}, Experiment. Math. \textbf{1}
  (1992), 275--305.

\bibitem[HS01]{HejhalStrombergsson2001}
D.~A. Hejhal and A.~Str\"{o}mbergsson, \emph{On quantum chaos and Maass
  waveforms of CM-type}, Found. Phys. \textbf{31} (2001), no.~3, 519--533.

\bibitem[Hun91]{Huntebrinker1991}
W.~Huntebrinker, \emph{Numerische Bestimmung von Eigenwerten des
  La\-pla\-ce-Operators auf hyperbolischen R\"{a}umen mit adaptiven
  Finite-Element-Methoden}, Bonner Mathematische Schriften \textbf{225}
  (1991).

\bibitem[Iwa95]{Iwaniec1995}
H.~Iwaniec, \emph{Introduction to the Spectral Theory of Automorphic
  Forms}, Revista Matem\'{a}tica Iberoamericana, 1995.

\bibitem[Kub73]{Kubota1973}
T.~Kubota, \emph{Elementary Theory of Eisenstein Series}, Kodansha,
  Tokyo and Halsted Press, 1973.

\bibitem[Maa49]{Maass1949}
H.~Maa{\ss}, \emph{\"{U}ber eine neue Art von nichtanalytischen
  automorphen Funktionen und die Bestimmung Dirichletscher Reihen
  durch Funktional\-glei\-chungen}, Math. Ann. \textbf{121} (1949),
  141--183.

\bibitem[Maa64]{Maass1964}
H.~Maa{\ss}, \emph{Lectures on Modular Functions of one Complex
  Variable}, Tata Institute of Fundamental Research, 1964, Springer, Revised
  1983.

\bibitem[Miy89]{Miyake1989}
T.~Miyake, \emph{Modular forms}, Springer, 1989.

\bibitem[MOS66]{MagnusOberhettingerSoni1966}
W.~Magnus, F.~Oberhettinger, and R.~P. Soni, \emph{Formulas and Theorems
  for the Special Functions of Mathematical Physics}, Springer, 1966.

\bibitem[Olv52]{Olver1952}
F.~W.~J. Olver, \emph{Some new asymptotic expansions for Bessel functions
  of large orders}, Proc. Camb. Phil. Soc. \textbf{48} (1952), 414--427.

\bibitem[Olv54]{Olver1954}
F.~W.~J. Olver, \emph{The asymptotic expansion of Bessel functions of large
  order}, Phil. Trans. A \textbf{247} (1954), 328--368.

\bibitem[Roe66]{Roelcke1966}
W.~Roelcke, \emph{Das Eigenwertproblem der automorphen Formen in der
  hyperbolischen Ebene}, Math. Ann. \textbf{167} (1966), 292--337 and
  \textbf{168} (1967), 261--324.

\bibitem[Sar95]{Sarnak1993}
P.~Sarnak, \emph{Arithmetic quantum chaos}, Israel Math. Conf. Proc. \textbf{8}
  (1995), 183--236.

\bibitem[Sch54]{Schoebe1954}
W.~Sch\"{o}be, \emph{Eine an die Nicholsonformel anschlie{\ss}ende
  asymptotische Entwicklung f\"{u}r Zylinderfunktionen}, Acta. Math.
  \textbf{92} (1954), 265--307.

\bibitem[Sch91]{Schmit1991}
C.~Schmit, \emph{Triangular billiards on the hyperbolic plane: Spectral
  properties}, preprint, IPNO/TH 91-68 (1991).

\bibitem[Sel56]{Selberg1956}
A.~Selberg, \emph{Harmonic analysis and discontinuous groups in weakly
  symmetric Riemannian spaces with applications to Dirichlet series}, J.
  Indian Math. Soc. \textbf{20} (1956), 47--87.

\bibitem[SS02]{SelanderStrombergsson2002}
B.~Selander and A.~Str\"{o}mbergsson, \emph{Sextic coverings of genus two
  which are branched at three points}, UUDM report 2002:16 (Uppsala 2002).

\bibitem[Sta84]{Stark1984}
H.~M. Stark, \emph{Fourier coefficients of Maass waveforms}, Modular
  Forms (R.~A. Rankin, ed.), Ellis Horwood, 1984, pp.~263--269.

\bibitem[Ste92]{Steil1992}
G.~Steil, \emph{\"{U}ber die Eigenwerte des Laplaceoperators und der
  Heckeoperatoren f\"{u}r $\operatorname{SL}(2,\mathbb{Z})$}, 1992,
  Diplomarbeit, Universit\"{a}t Hamburg, II. Institut f\"{u}r Theoretische
  Physik.

\bibitem[Ste94]{Steil1994}
G.~Steil, \emph{Eigenvalues of the Laplacian and of the Hecke operators
  for $\operatorname{PSL}(2,\mathbb{Z})$}, DESY report \textbf{94-028} (Hamburg
  1994).

\bibitem[Ter85]{Terras1985}
A.~Terras, \emph{Harmonic Analysis on Symmetric Spaces and
  Applications}, vol.~1, Springer, 1985.

\bibitem[Ven90]{Venkov1990}
A.~B. Venkov, \emph{Spectral Theory of Automorphic Functions and Its
  Applications}, Kluwer Academic Publishers, 1990.

\bibitem[Vig83]{Vigneras1983}
M.-F. Vign\'{e}ras, \emph{Quelques remarques sur la conjecture
  $\lambda_1\ge\frac{1}{4}$}, S\'{e}minaire de Th\'{e}orie des
  Nombres (M.-J. Bertin, ed.), Birkh\"{a}user, 1983, pp.~321--343.

\bibitem[Wat44]{Watson1944}
G.~N. Watson, \emph{A treatise on the theory of Bessel functions},
  Cambridge University Press, 1944.

\bibitem[Win88]{Winkler1988}
A.~M. Winkler, \emph{Cusp forms and Hecke groups},
  J. Reine Angew. Math. \textbf{386} (1988), 187--204.

\end{thebibliography}
\end{document}